\documentclass[runningheads]{llncs}
\usepackage[T1]{fontenc}
\usepackage{graphicx}
\usepackage{xcolor}

\usepackage{tabularx}
\usepackage{booktabs}
\usepackage{amsmath}
\usepackage{float}
\usepackage{multirow}
\usepackage{diagbox}
\usepackage{makecell}
\usepackage{subcaption}

\usepackage[hidelinks]{hyperref}
\usepackage[capitalize]{cleveref}
\begin{document}
\title{Knowledge Synthesis Graph:\\
An LLM-Based Approach for Modeling Student Collaborative Discourse}
\titlerunning{An LLM-Based Approach for Modeling Student Collaborative Discourse}
%
\author{Bo Shui\inst{1}\orcidID{0009-0001-0981-9608} \and
Xinran Zhu\inst{1}\orcidID{0000-0003-0064-4861}}
\authorrunning{B. Shui and X. Zhu}
%
\institute{University of Illinois Urbana-Champaign, Champaign IL 61820, USA \\
\email{\{boshui2,xrzhu\}@illinois.edu}}
\maketitle              

\begin{abstract}
Asynchronous, text-based discourse---such as students’ posts in discussion forums---is widely used to support collaborative learning. However, the distributed and evolving nature of such discourse often makes it difficult to see how ideas connect, develop, and build on one another over time. As a result, learners may struggle to recognize relationships among ideas---a process that is critical for idea advancement in productive collaborative discourse. To address this challenge, we explore how large language models (LLMs) can provide representational guidance by modeling student discourse as a Knowledge Synthesis Graph (KSG). The KSG identifies ideas from student discourse and visualizes their epistemic relationships, externalizing the current state of collaborative knowledge in a form that can support further inquiry and idea advancement. In this study, we present the design of the KSG and evaluate the LLM-based approach for constructing KSGs from authentic student discourse data. Through multi-round human-expert coding and prompt iteration, our results demonstrate the feasibility of using our approach to construct reliable KSGs across different models. This work provides a technical foundation for modeling collaborative discourse with LLMs and offers pedagogical implications for augmenting complex knowledge work in collaborative learning environments.

\end{abstract}
\section{Introduction}
Asynchronous, text-based discourse—such as students' posts in discussion forums—is widely used to support collaborative learning in technology-mediated environments. The productivity of such discourse depends on learners’ ability to collaboratively connect, build on, and advance their ideas over time~\cite{Chan2024CSCL,zhu2026advancing}. However, in asynchronous settings, ideas are often distributed across platforms, threads, and time, making these processes difficult to sustain~\cite{suthers2010framework}. As a result, learners may struggle to see connections between ideas or understand how individual ideas contribute to larger lines of inquiry~\cite{zheng2023automatic}. This highlights the need for supporting structures that help make student ideas more visible, organized, and actionable for further learning.

One common form of such support is the use of \textit{representational guidance} for student ideas~\cite{suthers2001towards}. For example, instructors may create summaries or concept maps to surface key ideas from student discourse and steer subsequent conversations. Such external representations—such as diagrams or visual maps—can shape how people think and collaborate by inviting further inquiry and collaborative advancement~\cite{hennessy2011role}. However, creating such guidance can be labor-intensive, especially when student contributions are numerous and continuously evolving.

Recent advances in large language models (LLMs) offer new possibilities for generating such guidance by analyzing discourse at scale, surfacing salient ideas, and modeling relationships among them. At the same time, realizing this potential requires careful attention to two key tensions. First, LLMs tend to prioritize dominant statistical patterns, which can oversimplify nuanced human input by overlooking minority perspectives and flattening productive contradictions that drive deep learning~\cite{kapur2008productive,bender2021dangers}. Second, there are growing concerns that LLM-generated content may undermine human agency by positioning learners as passive recipients rather than active co-creator of knowledge. 

In response, we explore how LLMs can be used to generate representational guidance for idea advancement while preserving epistemic nuance and learner agency. We introduce an approach for constructing a \textit{Knowledge Synthesis Graph} (KSG) that captures and visualizes the evolving state of student thinking through collaborative discourse. Grounded in computer-supported collaborative learning (CSCL) literature, we conceptualize student discourse as a \textit{knowledge synthesis} process---integrating diverse ideas to foster conceptual innovation, generate new knowledge, and support creative problem-solving~\cite{yuan2022cross,zhu2026advancing}. Within this context, the KSG functions as an \textit{intermediate synthesis artifact}: a networked graph that identifies key ideas, surfaces emerging themes, and models relationships among them to scaffold ongoing knowledge building. 

In this paper, we present the design and technical pipeline used to construct KSGs. We focus specifically on one form of discourse: social annotation—a common practice where students highlight and comment on shared readings using platforms like Perusall. We then evaluate the pipeline using authentic student discourse data from a social annotation activity. Specifically, we ask: \textit{To what extent can our approach generate accurate and reliable intermediate synthesis artifacts that capture the epistemic content of student discourse?} This work provides a proof of concept for applying LLMs to support collaborative learning in ways grounded in learning sciences. Our contributions are twofold. First, we introduce a representational approach that treats student discourse as a knowledge synthesis process. Second, we present and evaluate an LLM-based pipeline for KSG construction, aiming to preserve the nuance and complexity of student thinking. This approach also lays the foundation for a learning platform under development by our team, where learners iteratively refine synthesis graphs in collaboration with an AI partner.

\vspace{-0.8em}
\section{Design of the Knowledge Synthesis Graph}
\label{sec:design}
\begin{figure}[htbp]
    \vspace{-1em}
    \centering
    \includegraphics[width=0.5\textwidth]{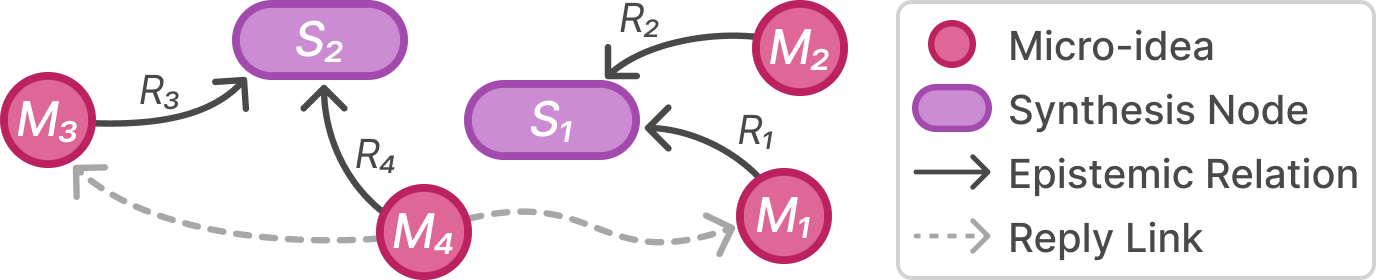}
    \caption{An illustrative example of the Knowledge Synthesis Graph.} 
    \label{fig:graph}
    \vspace{-2em}
\end{figure}

\vspace{-0.5em}
We propose a generic representational structure for the KSG consisting of two types of nodes---\textit{Micro-idea} and \textit{Synthesis Node}---and the \textit{Epistemic Relation} between them. In this study, we use social annotation as a concrete example to illustrate construction of the KSG. The representational design is informed by prior work on discourse graphs, concept mapping, and scholarly knowledge representation~\cite{chan2024steps,novak1984learning,de2009hypotheses}. Figure~\ref{fig:graph} illustrates the structure of a KSG.

Micro-ideas are fine-grained epistemic units derived from student social annotations. Because social annotations are often short, context-dependent, and conversational, LLMs struggle to reliably capture the epistemic, contextual, and rhetorical intent embedded in them. To address this challenge, we preprocess annotations into Micro-ideas---concise, standalone statements that explicitly express the core epistemic intent of the original contribution. This step aims to enable more accurate and consistent LLM outputs.

Synthesis Nodes are higher-level ideas that represent connections across multiple student annotations during the synthesis process. They capture broader conceptual structures or patterns within the discourse, while remaining open to further development. In the KSG, these nodes are not generated from student annotations directly. Instead, they are derived from key concepts and arguments in the course readings. This design provides a stable, shared reference point that helps students relate their own annotations to central ideas in the course. As students engage with the KSG by linking, revising, and extending ideas, Synthesis Nodes can also evolve, supporting deeper understanding and more integrated knowledge work over time.

Epistemic Relations describe how Micro-ideas connect to Synthesis Nodes in ways that support future synthesis. These relations are forward-looking, indicating how a Micro-idea might contribute to the evolvement of a Synthesis Node. By making these connections explicit, the KSG shows how student ideas from annotations can drive deeper synthesis in ongoing learning activities.

\vspace{-0.5em}
\section{Pipeline for Knowledge Synthesis Graph Construction}
\label{sec:pipeline}
\begin{figure}[ht]
    \vspace{-1em}
    \includegraphics[width=\textwidth]{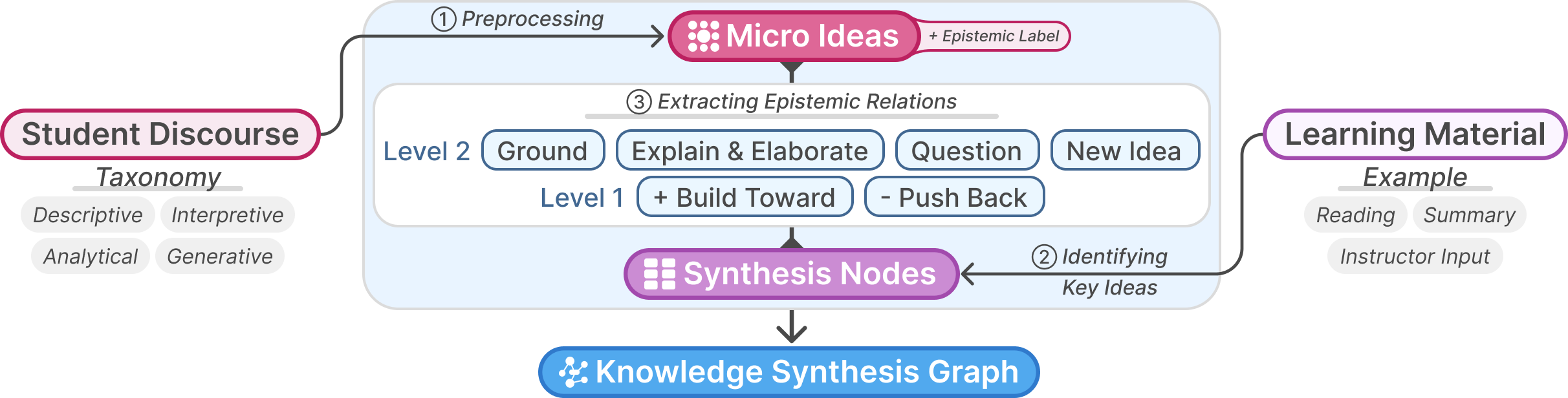}
    \caption{Technical pipeline for KSG construction.} 
    \label{fig:framework}
    \vspace{-1em}
\end{figure}

\vspace{-0.5em}
Our approach is implemented as a three-stage pipeline, incrementally processing discourse into a KSG leveraging LLMs, as illustrated in Fig~\ref{fig:framework}. Each stage was developed through iterative prompt engineering. 

\textbf{Stage 1: Preprocessing student discourse to extract Micro-ideas for context-aware LLM processing.} 
In this stage, LLMs follow a Contextualize $\rightarrow$ Filter $\rightarrow$ Assign pattern to transform raw annotations into Micro-ideas. Each prompt includes the original annotation, a quoted passage from the course reading, and any associated reply context. The model filters out non-substantive discourses (e.g., "I agree") and rewrites the remaining content into a standalone, concise statement that preserve the core epistemic meaning. 

Simultaneously, the model assigns a label identifying the epistemic function of each Micro-idea. We developed the coding scheme through an iterative process involving two researchers in Learning Sciences and HCI. The final scheme includes four types of labels---descriptive, interpretive, analytical, and generative---as shown in Table~\ref{tab:micro_idea_types}. They reflect varying levels of epistemic engagement and signal how a Micro-idea may contribute to subsequent knowledge synthesis.

\begin{table}[htb]
    \vspace{-2.5em}
    \centering
    \setlength{\intextsep}{0pt} 
    \setlength{\abovetopsep}{0pt} 
    \setlength{\belowbottomsep}{0pt}
    \caption{Taxonomy of Micro-idea Labels}
    \label{tab:micro_idea_types}
    \resizebox{0.9\textwidth}{!}{
    \begin{tabularx}{\textwidth}{l @{\hspace{0.6em}} X}
    \toprule
    \textbf{Type} & \textbf{Epistemic Function} \\ \midrule
    \textbf{Descriptive} & States perspectives from  the text to establish shared understanding. \\ 
    \textbf{Interpretive} & Moves beyond restating the text to explains meaning, significance, or implications with some elaboration. \\ 
    \textbf{Analytical} & Breaks down relationships or mechanisms; judges strength, limitations, quality, or assumptions to helps build structure for bigger claims and surfaces critical stance needed for advanced synthesis. \\ 
    \textbf{Generative} & Proposes a new idea, hypothesis, or design implication to support idea advancement and synthesis creation. \\ \bottomrule
    \end{tabularx}
    }
    \vspace{-1.5em}
\end{table}



\textbf{Stage 2: Identifying Synthesis Nodes.} In this stage, the LLMs extracts key ideas from the course readings to generate candidate Synthesis Nodes. Using the full text of the reading along with a summary and instructor-provided prompts, the model generates core concepts and arguments that represent central themes in the material. 


\textbf{Stage 3: Extracting Epistemic Relations to link Micro-ideas to Synthesis Nodes.} In the final stage, LLMs identify Epistemic Relations that link Micro-ideas to relevant Synthesis Nodes. 

Informed by prior work on discourse graphs~\cite{chan2020knowledge,morabito2021managing}, argumentation theory~\cite{toulmin2003uses}, and rhetorical moves in educational contexts~\cite{knight2017towards}, we developed a two-level coding scheme to describe the Epistemic Relations. As shown in Table~\ref{tab:epistemic_relations}, Level 1 captures the epistemic stance of a Micro-idea relative to a Synthesis Node (i.e., whether it builds toward or pushes back against the Synthesis Node); Level 2 further specifies the epistemic function of the Micro-idea within each stance using four categories. Together, this scheme represents both the orientation and the functional role of student contributions in relation to synthesis. 

LLMs receive the Micro-ideas and Synthesis Nodes generated in previous stages and use this coding scheme to assign Epistemic Relations. When a Micro-ideas cannot be confidently linked, it is assigned to an "uncategorized" node, inviting students to manually refine the graph for ongoing knowledge synthesis.

\begin{table}[htbp]
    \vspace{-2em}
    \centering
    \caption{Epistemic Relation Coding Scheme}
    \label{tab:epistemic_relations}
    \resizebox{0.9\textwidth}{!}{
    \begin{tabularx}{\textwidth}{l @{\hspace{0.6em}} X @{\hspace{0.6em}} X}
    \toprule
     & \textbf{Build toward ($+$)} & \textbf{Push back ($-$)} \\ 
    \midrule
    \textbf{Ground} & Provides validating evidence or examples to clarify a claim. & Provides counter-examples or edge cases to identify limitations. \\
    \makecell[tl]{\textbf{Explain \&} \\ \textbf{Elaborate}} & Unpacks meaning or adds details to extend current ideas. & Deconstructs to reveal assumptions or internal contradictions. \\
    \textbf{New Idea} & Introduces concepts that connect other parts of synthesis. & Introduces competing ideas that shift the focus away. \\
    \textbf{Question} & Probes for more detail to help refine the synthesis. & Raises critics about validity, bias, or logic to prompt revision. \\ 
    \bottomrule
    \end{tabularx}
    }
    \vspace{-1em}
\end{table}
\vspace{-0.5em}
\section{Evaluation}
\vspace{-0.5em}

This section presents a preliminary evaluation of the proposed approach, examining the reliability and feasibility of Micro-idea extraction, Synthesis Node generation, and Epistemic Relation linking. The evaluation was informed by human-expert review and iterative prompt engineering. We analyzed a dataset of 42 annotations from a graduate-level instructional design course. Evaluation methods included descriptive analyses of expert analytic memos and model outputs, along with quantitative metrics assessing agreement and consistency. These analyses aimed to determine whether the framework produces stable and meaningful representations of student discourse. 

\vspace{-0.5em}
\subsection{Accuracy of Micro-idea Extraction}
\vspace{-0.5em}
We began by evaluating the accuracy of Micro-idea extraction, using the epistemic labels as a proxy for how well the model captured the intent and substance of annotations. This evaluation was closely tied to iterative prompt refinement: expert coders reviewed outputs and provided feedback that informed successive revisions to the prompts used to guide the model. The process unfolded over three prompt versions, each designed to address observed limitations:

\vspace{-0.3em}
\begin{itemize}
    \item $P_{\text{base}}$ introduced a basic four-category labeling system;
    \item $P_1$ added more detailed linguistic cues and decision logic;
    \item $P_2$ further clarified epistemic features and improved guidance for interpreting question-based annotations.
\end{itemize}
\vspace{-0.3em}

\begin{figure}[htbp]
    \centering
    \vspace{-1em}
    \includegraphics[width=0.4\textwidth]{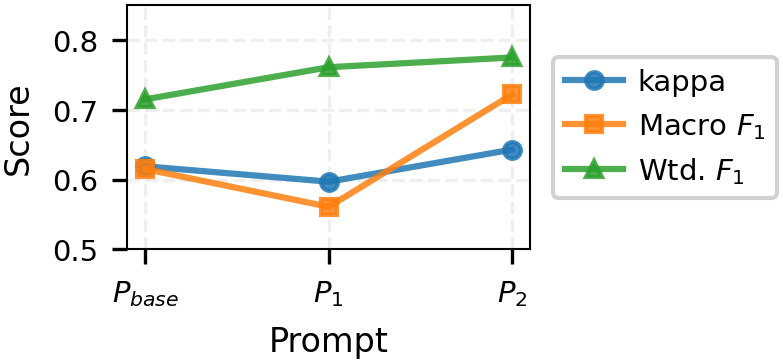}
    \caption{Agreement of Micro-idea labeling in stage 1.} 
    \label{fig:stage_1_agreement}
    \vspace{-2em}
\end{figure}

As shown in Fig~\ref{fig:stage_1_agreement}, model performance (using GPT-4o) was assessed at each stage by comparing its outputs to expert-coded labels. While $P_{\text{base}}$ achieved moderate initial agreement ($\kappa = 0.619$), performance dipped with $P_1$ as added complexity introduced new ambiguity (Macro $F_1=0.561$). In contrast, $P_2$ showed clear improvement across all metrics: $\kappa = 0.643$ (inter-rater reliability), Macro $F_1 = 0.722$ (balanced agreement across types), and Weighted $F_1 = 0.775$ (overall accuracy). The increase in Macro $F_1$—which balances performance across epistemic label types—suggests that making epistemic features more explicit helped the model better detect less frequent but pedagogically meaningful contributions.


\vspace{-0.5em}
\subsection{Reliability of Synthesis Nodes Generation}
\vspace{-0.5em}

We next evaluated the reliability of Synthesis Nodes generation by analyzing model outputs across different prompt configurations, using descriptive analysis of expert analytic memos collected during prompt iteration. Expert evaluation centered on two primary criteria: how well the generated nodes aligned with instructor-provided prompts and how coherently they reflected the main ideas of the reading. Results revealed a tradeoff between the amount and structure of context in the prompt and the quality of generated content. Limited context (e.g., abstract and introduction only) often led to vague or surface-level nodes lacking epistemic depth. Conversely, providing extensive, unstructured full text increased the risk of hallucination and reduced topical focus. We found that including a high-level reading summary helped the models generate more concise and coherent nodes while preserving core ideas. Additionally, incorporating instructor-provided prompts further improved alignment with pedagogical intent. Overall, these analyses suggest that combining a high-level summary with instructor input offers the most effective configuration for generating reliable and pedagogically meaningful Synthesis Nodes.

\vspace{-0.5em}
\subsection{Cross-model Consistency of Epistemic Relation Linking}
\vspace{-0.5em}

Building on results from the previous stages, we evaluated the model's ability to link Micro-ideas to appropriate Synthesis Nodes by identifying both the correct Epistemic Relation and the target node. Given the open-ended nature of relations and the inherent subjectivity of knowledge synthesis, we chose not to evaluate model performance based on alignment with human-coded labels. Instead, the evaluation presented in this study focused on consistency across model outputs. We tested four prompt configurations, each reflecting a different coding scheme:

\vspace{-0.4em}
\begin{itemize}
    \item $P_{\text{base}}$: 4-category basic argumentative stances (support, challenge, exemplify, question);
    \item $P_1$: 6-category for greater granularity (evidence, explain, challenge, qualify, summarize, extend);
    \item $P_2$: 3-category for reducing confusion (support, critique, reflect);
    \item $P_3$: the proposed two-level coding scheme (see Table~\ref{tab:epistemic_relations}).
\end{itemize}
\vspace{-0.2em}

\begin{figure}[htbp]
    \vspace{-1em}
    \includegraphics[width=0.9\textwidth]{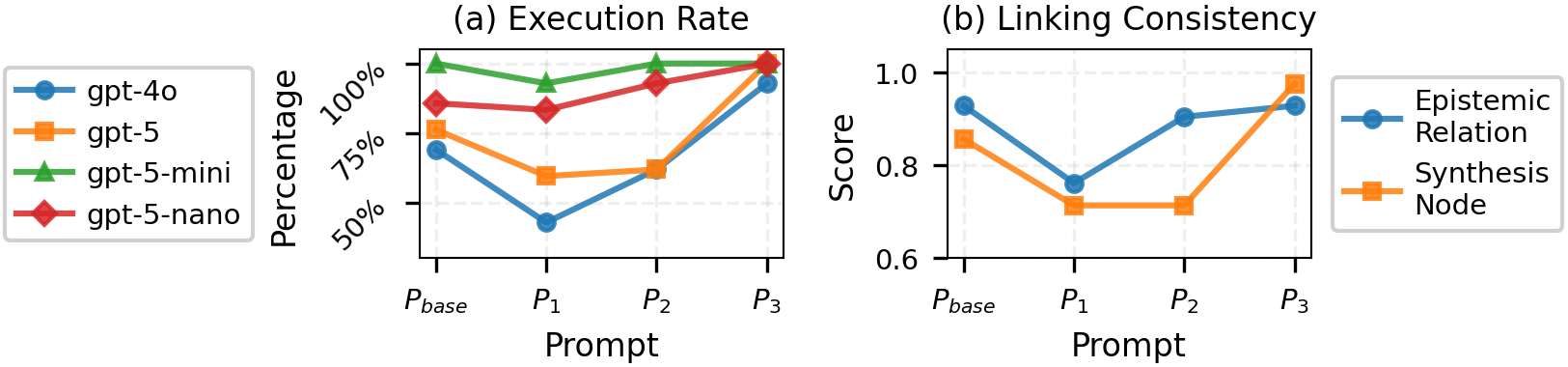}
    \caption{Cross-model execution rate and linking consistency in stage 3.} 
    \label{fig:stage_3_analysis}
    \vspace{-1em}
\end{figure}

We evaluated two metrics across four LLMs (GPT-4o and GPT-5 series): execution rate (the percentage of valid output) and linking consistency (the frequency of two or more models producing identical output). As shown in Fig~\ref{fig:stage_3_analysis}, $P_1$ reduced both execution and consistency, suggesting that increasing granularity without structural guidance hinder performance. The simplified $P_2$ restored execution stability but did not improve consistency. In contrast, $P_3$ resulted in improvements on both metrics, indicating that this structure provides a more stable and interpretable foundation to transforming discourse into reliable graphs.




\vspace{-1em}
\section{Discussion}


This paper presented an early-stage, LLM-based approach for constructing Knowledge Synthesis Graphs to visualize and scaffold student collaborative discourse. Through iterative design, evaluation, and prompt engineering, we explored how intermediate synthesis artifacts---Micro-ideas, Synthesis Nodes, and Epistemic Relations---can be generated to represent the emerging state of collaborative knowledge construction. Our findings highlight both the promise and the practical challenges of applying LLMs in educational settings, and underscore the need for thoughtful prompt configuration and intentional alignment with pedagogical goals to ensure that model outputs meaningfully support learning. A core insight from our work is the need to rethink how Generative AI systems are evaluated in education. Rather than optimizing for accuracy against a single ground truth, future approaches should account for context, learner interaction, and pedagogical intent. This calls for expanding evaluation frameworks to capture how AI systems shape, mediate, and participate in the human learning process---not just what they produce.

Building on this early-stage work, we identify three key directions for future development. First, we plan to continue prompt iteration and evaluation through classroom implementation and user studies to assess how students interact with and benefit from KSG in authentic settings. Second, the challenge in generating epistemic links points to the need for more structured reasoning strategies. Future work will explore advanced prompting techniques, such as chain-of-thought reasoning~\cite{wei2022chain} and verifier models~\cite{yang2025llm2}, to improve both the reliability and interpretability of relational linking. Third, we envision the Knowledge Synthesis Graph not as a static output, but as a dynamic, improvable artifact that evolves alongside students’ collaborative knowledge work. This opens up opportunities to design interactive systems where AI-generated graphs serve as prompts for further elaboration, and student contributions iteratively refine and reshape the graph—paving the way for responsive, learner-centered human–AI collaboration.

\begin{credits}
\subsubsection{\ackname}
The authors would like to thank Liam Magee and Ishita Asnani for their valuable early contributions to the design of the Knowledge Synthesis Graph.

\end{credits}
%
%
\vspace{-0.5em}
\bibliographystyle{splncs04}
\bibliography{mybib}
%




\end{document}